\documentclass[journal]{IEEEtran}

\usepackage{contour}
\usepackage{float}
\usepackage{graphicx,xcolor}
\usepackage{cite}
\usepackage{comment}
\usepackage{amsmath}
\usepackage{amssymb}
\usepackage[outdir=./]{epstopdf}
\usepackage{multirow}
\usepackage[utf8]{inputenc}
\usepackage[T1]{fontenc}
\usepackage{tikz}
\usetikzlibrary{positioning}
\usetikzlibrary{shapes,arrows,fit}
\usetikzlibrary{shapes.gates.logic.US}
\usetikzlibrary{spy}              
\usetikzlibrary{calc}             
\usepackage{soul}
\usepackage[normalem]{ulem}
\usepackage[numbers,square,sort&compress]{natbib}
\usepackage{glossaries}                
\usepackage{algorithm}                 
\usepackage{algorithmic}               
\usepackage{amsfonts}                  
\usepackage{bm}                        
\usepackage{booktabs}                  
\usepackage{multirow}                  
\usepackage{pgf}
\makeatletter                          
\let\pgfmathModX=\pgfmathMod@          
\usepackage{pgfplots}
\let\pgfmathMod@=\pgfmathModX          
\makeatother
\usepackage{pgfplotstable}             
\usepackage[caption=false]{subfig}     
\usepackage{todonotes}
\usepackage{marginnote}
\makeatletter

\renewcommand{\@todonotes@backgroundcolor}{white}
\renewcommand{\@todonotes@textwidth}{2.1\marginparwidth}
\renewcommand{\@todonotes@drawMarginNoteWithLine}{%
\begin{tikzpicture}[remember picture, overlay, baseline=-0.75ex]%
    \node [coordinate] (inText) {};%
\end{tikzpicture}%
\marginnote[{
    \@todonotes@drawMarginNote%
    \@todonotes@drawLineToLeftMargin%
}]{
    \@todonotes@drawMarginNote%
    \@todonotes@drawLineToRightMargin%
}%
}
\makeatother

\makeatletter
\if@todonotes@disabled

\else

\fi
\makeatother

\usepackage[hidelinks]{hyperref}
\hypersetup{
    colorlinks,
    linkcolor={red!50!black},
    citecolor={blue!50!black},
    urlcolor={blue!80!black}
}

\definecolor{darkGrey}{RGB}{90,90,90}
\definecolor{pastelLucio}{RGB}{247,233,134}
\definecolor{blueLucio}{RGB}{223,233,249}
\definecolor{pinkLucio}{RGB}{248,233,232}
\definecolor{pastelYellow}{RGB}{255,251,241}

\newcommand\kaWuline{\bgroup\markoverwith{\textcolor[rgb]{0.12,0.74,0.36}{\rule[-0.4ex]{2pt}{0.4pt}}}\ULon}
\newcommand\vlWuline{\bgroup\markoverwith{\textcolor[rgb]{0.74,0.12,0.36}{\rule[-0.4ex]{2pt}{0.4pt}}}\ULon}
\newcommand\alWuline{\bgroup\markoverwith{\textcolor[rgb]{.6,0.5,0}{\rule[-0.4ex]{2pt}{0.7pt}}}\ULon}
\newcommand\cacWuline{\bgroup\markoverwith{\textcolor{orange}{\rule[-0.5ex]{2pt}{0.4pt}}}\ULon}

\usepackage{marginfix}

\tikzset{%
  block/.style    = {draw, thick, rectangle, minimum height = 2em,
    minimum width = 2em},
  sum/.style      = {draw, circle, semithick, node distance = 1.0cm}, 
  cross/.style={path picture={ 
      \draw[black]
      (path picture bounding box.south east) -- (path picture bounding box.north west) (path picture bounding box.south west) -- (path picture bounding box.north east);
    }},
  input/.style    = {coordinate}, 
  output/.style   = {coordinate} 
}

\tikzstyle{process} = [rectangle, minimum width=1.3cm, minimum height=1.3cm, text centered, draw=black, semithick, fill=blueLucio]
\tikzstyle{process2} = [rectangle, minimum width=1.8cm, minimum height=1.0cm, text centered, draw=black, semithick, fill=pinkLucio]
\tikzstyle{cloud} = [node distance=1.1cm,   minimum height=0em]
\tikzset{font={\fontsize{8pt}{12}\selectfont}}
\tikzstyle{arrow} = [thick,->,>=stealth]

\pgfplotstableset{
  col sep=comma,                 
  every head row/.style={        
    before row=\toprule,   %
    after row=\midrule     %
  },                             %
  every last row/.style={        %
    after row=\bottomrule  %
  },                             %
  fixed,                         
  fixed zerofill,                %
  precision=2,                   %
}

\def\showdetail#1#2#3#4#5#6{%
  \node[image,#2] (#1) {\includegraphics[width=#6]{figure/#1}};
  \path (#1.south west) ++(#3) coordinate (#1 detail position);
  \spy on (#1 detail position) in node [above=of #1];
  \node[below=1mm of #1] (#1 detail) {#4};
}

\def\showdetailall#1#2#3#4#5#6#7{%
  \def\imw{#3} 
  \begin{tikzpicture}[
    spy using outlines={%
      rectangle,
      magnification = #6,
      width = #4,
      height = #5,
      connect spies,
      very thick,
      red,
    },
    node distance = #7,
    image/.style = {anchor=south west, inner sep=0},
    ]
    \def\printpsnr{\pgfmathprintnumber{\pgfplotsretval}dB}
    \showdetail{#1}{}{#2}{Ground truth}{#5}{#3}
    \pgfplotstableread{data/psnr_#1_sigma50.csv}\psnrtable
    \pgfplotstablegetelem{0}{noisy}\of\psnrtable
    \showdetail{#1_sigma50}{right=of #1}{#2}{Noisy \printpsnr}{#5}{#3}
    \pgfplotstablegetelem{0}{bm3d}\of\psnrtable
    \showdetail{#1_sigma50_bm3d}{right=of #1_sigma50}{#2}{BM3D \cite{BM3D-TIP} \printpsnr}{#5}{#3}
    \pgfplotstablegetelem{0}{wdncnn}\of\psnrtable
    \showdetail{#1_sigma50_wdncnn}{right=of #1_sigma50_bm3d}{#2}{\wavdncnn\ \cite{WavDnCNN} \printpsnr}{#5}{#3}
    \pgfplotstablegetelem{0}{nn3dwdncnnbmcnn}\of\psnrtable
    \showdetail{#1_sigma50_wdncnn_bmcnnf}{right=of #1_sigma50_wdncnn}{#2}{\thingcombo{\wavdncnn} \printpsnr}{#5}{#3}
  \end{tikzpicture}
}

\def\wavdncnn{\text{WDnCNN}}
\def\thing{{NN3D}}
\def\qWiener{q}
\newcommand{\thingsuffix}{\thing}
\newcommand{\thingcombo}[1]{\thingsuffix\big(#1\big)}

\def\cnnTrainStd{\varsigma}
\def\cnnTrainStdSet{\Sigma}
\newcommand{\mybreve}[1]{#1}
\newacronym{awgn}{AWGN}{additive white Gaussian noise}
\newacronym{bm}{BM}{Block matching}
\newacronym{bm3d}{BM3D}{Block Matching 3D}
\newacronym{bmcnn}{BMCNN}{block matching convolutional neural network}
\newacronym{cnn}{CNN}{convolutional neural network}
\newacronym{cnnf}{CNNF}{\gls{cnn}-based filter}
\newacronym{gpu}{GPU}{graphics processing unit}
\newacronym{nlf}{NLF}{nonlocal filter}
\newacronym{nlm}{NLM}{Nonlocal Means}
\newacronym{wsd}{WSD}{Wiener filter in Similarity Domain}
\newacronym{psnr}{PSNR}{peak signal to noise ratio}
\newacronym{ssim}{SSIM}{structural similarity index measure}

\newcommand{\rev}[2]{#2}                         

\newcommand{\revv}[2]{#2}                         

\begin{document}

\title{Nonlocality-Reinforced Convolutional Neural Networks for Image Denoising}
\author{Cristóvão~Cruz, 
  Alessandro~Foi,~\IEEEmembership{Senior Member,~IEEE,}
  Vladimir~Katkovnik, 
   and~Karen~Egiazarian,~\IEEEmembership{Fellow,~IEEE}
  \thanks{This work is supported by the Academy of Finland (projects no.~287150, 
    2015-2019, and no.~310779, 2017-2021) and European Union's H2020 Framework Programme (H2020-MSCA-ITN-2014) 
under grant agreement no.~642685 MacSeNet. The authors are with Tampere University of Technology, Finland, and with Noiseless Imaging Ltd, Finland. e-mail: cristovao@noiselessimaging.com}} 

\markboth{IEEE SPL}%
{Shell \MakeLowercase{\textit{et al.}}: Bare Demo of IEEEtran.cls for IEEE Journals}
%


\maketitle

\begin{abstract}
  We introduce a paradigm for nonlocal sparsity reinforced deep
  convolutional neural network denoising. It is a combination of a
  local multiscale denoising by a convolutional neural network (CNN)
  based denoiser and a nonlocal denoising based on a \gls{nlf}
  exploiting the mutual similarities between groups of patches.  CNN
  models are leveraged with noise levels that progressively decrease
  at every iteration of our framework, while their output is
  regularized by a nonlocal prior implicit within the
  \gls{nlf}. Unlike complicated neural networks that embed the
  nonlocality prior within the layers of the network, our framework is
  modular, it uses standard pre-trained CNNs together with standard
  nonlocal filters.  An instance of the proposed framework, called
  \thing, is evaluated over large grayscale image datasets showing
  state-of-the-art performance.
\end{abstract}


\glsresetall

\section{Introduction}

\IEEEPARstart{I}mage denoising, one of the most important problems of
image processing and computer vision, aims at estimating an unknown
image from its noisy observation. Image denoising plays a crucial role
at various stages of image processing and can be used as a replacement
of explicit image priors:

\noindent \emph{in pre-processing}, to enhance the output quality and
performance of the subsequent image-processing or computer-vision
tasks, such as demosaicing, sharpening, compression, object
segmentation, classification, and recognition
\cite{milanfar_2013_tour};

\noindent \emph{in post-processing: } to suppress compression
artifacts, such as blocking and ringing ~\cite{FoiSADCT};

\noindent \emph{as plug\&play} filter: i.e. as an implicit
regularization prior used in various inverse-imaging applications
\cite{ChanPlug} and in end-to-end optimized computational imaging
systems \cite{FlexISP}.

\contourlength{2.5pt}
\contournumber{60} 
\begin{figure}[t]
  \centering
  \begin{tikzpicture}[node distance = 0.7cm, auto]
  \linespread{0.65}
    \node (in1) [cloud, align=center]
    {$z$};
    \node (in2) [cloud, align=center, below=22.5mm of in1] {$\hat{y}_{k-1}$\hspace*{-0.05cm}};
    \node (out2) [cloud, align=center, below right=22.5mm and 7.3cm of in1] {$\hat{y}_{k}$};
    \node (wsum) [sum, inner sep=0.5pt ,below right=10mm and 9mm of in1] {$\boldsymbol{+}$};
    \node (amp1) [draw, shape border rotate=180, regular polygon, regular polygon sides=3, semithick, minimum size=4mm, outer sep=-0.1pt,inner sep=0pt, above=4mm of wsum] {};
    \node (amp2) [draw, shape border rotate=0, semithick,regular polygon, regular polygon sides=3, minimum size=4mm, outer sep=-0.1pt, inner sep=0pt, below=4.3mm of wsum] {};
    \node (amp1ll) [cloud, align=center, below left=0.5mm and 2.5mm of amp1] {};
    \node (amp1ur) [cloud, align=center, above right=-0.2mm and 3mm of amp1] {};
    \node (amp2ll) [cloud, align=center, below left=-0.3mm and 2.5mm of amp2] {};
    \node (amp2ur) [cloud, align=center, above right=0.98mm and 3mm of amp2] {};
    \node (amp1llt) [align=right, above left=-0.7mm and -5mm of amp1ll] {$_{\lambda_k}$};
    \node (amp2llt) [align=right, above left=-0.4mm and -5.5mm of amp2ll] {$_{1-\lambda_k}$};
    
    \node (CNN) [process, right=10mm of wsum, align=center] {CNNF
    };
    \node (nlf) [process, right=11mm of CNN ,align=center] {{NLF}
    };
    \draw [arrow, thin] (amp1ll) -- (amp1ur);
    \draw [arrow, thin] (amp2ll) -- (amp2ur);
    \draw [arrow] (wsum) -- node[anchor=south] {\hspace*{-3mm}$\bar z_{k}$} (CNN.west) node[rotate=-90, 
    above,
    inner sep=1.5pt]{\scalebox{.76}{\tiny \textsf{NOISY}}};
    \draw [arrow] (CNN.east) node[rotate=90, 
    above,
    inner sep=1.5pt]{\scalebox{.76}{\tiny \textsf{DENOISED}}} -- node(zTilde)[anchor=south] {{\hspace*{-0.0mm}{$\tilde y_{k}$}}} (nlf.west) node[rotate=-90,    above,
    inner sep=1.5pt]{\scalebox{.76}{\tiny \textsf{NOISY}}};
    \draw [arrow] (amp1)   --  (wsum);
    \draw [arrow] (nlf.east) node[rotate=90, 
    above,
    inner sep=1.5pt]{\scalebox{.76}{\tiny \textsf{\ DENOISED}}} --  ++(0.6,-0.0) |- (out2.west);
    \draw [arrow] (amp2.north)  --  (wsum.south);
    \draw [thick] (in1.east)  -| (amp1.north);
    \draw [thick] (in2.east)  -|  (amp2.south);

    \node (par1) [cloud, align=center, below=4.8mm of CNN] {$\lambda^2_k\sigma^2$\ }; \draw [arrow] (par1.north) -- (CNN.south) node[above, inner sep=1.5pt]{\scalebox{.76}{\tiny \textsf{VARIANCE}}};
    \node (par2) [cloud, align=center, below=5.5mm of nlf.270] {\ $\tau_k$}; \draw [arrow] (par2.north) -- (nlf.270) node[above, inner sep=1.5pt]{\scalebox{.76}{\tiny \textsf{THRESHOLD\ \,}}}; 
  \end{tikzpicture}\vspace*{-0.2mm}
  \caption{Flowchart of the generic $k$-th iteration of the proposed 
    framework. 
    The noisy image $z$ is combined with a previous estimate $\hat{y}_{k-1}$ of the clean image $y$ as the convex combination $\bar{z}_k=\lambda_k z + \left(1-\lambda_k\right)\hat{y}_{k-1}$.
    First, $\bar{z}_k$ is filtered by a convolutional neural network filter (CNNF); then the CNNF output $\tilde{y}_k$ is processed by a Nonlocal filter (\gls{nlf}), whose 
output 
constitutes the new estimate $\hat{y}_{k}$. 
    The method iterates 
    with $1\!=\!\lambda_{1}\!>\!\dots\!>\lambda_{k-1}\!>\!\lambda_{k}\!>\!\dots\!>\!0$. \rev{The dashed line encloses the core elements for CNN denoising reinforced by nonlocal self-similarity; block matching determines the nonlocal self-similarity embedded in the NLF and is performed only at the first iteration.}{}
  } 
  \label{flowChart}	
\end{figure}
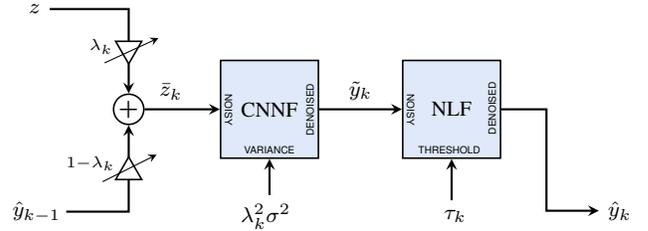

Recently, image denoising received a new boost of interest through the
application of advanced machine-learning methods, particularly deep
\glspl{cnn} \cite{zhangCvpr17}. The current most effective image
denoising methods can be roughly categorized into \emph{nonlocal
  filters (\glspl{nlf})} (e.g., NLBayes \cite{NL-Bayes}, \gls{bm3d}
\cite{BM3D-TIP,BM3DSAPCA}, and WNNM \cite{WNNM}) or
\emph{\glspl{cnnf}} (e.g., TNRD \cite{TNRD}, DnCNN \cite{DnCNN}).
Their respective advantages and drawbacks are:

\noindent{\it Advantages of \glspl{nlf}: } Joint collaborative
processing of mutually similar image patches, resulting in superior
noise removal where the image exhibits strong self-similarity, such on
edges or on regular texture.

\noindent{\it Drawbacks of  \glspl{nlf}: }
Inferior performance on pseudo-random textures or singular features,
i.e. where the image has weak self-similarity.

This drawback of \glspl{nlf} is partly overcome through the use of
external dictionaries \cite{mosseri2013combining}, and multi-scale
\cite{ebrahimi2008examining,zontak2013separating}, as well as
multi-stage iterative approaches \cite{WNNM}, which improve the
denoising quality by refining the matching of mutually similar blocks
and their shrinkage.

\noindent{\it Advantages of \glspl{cnnf}: } a) ability to learn and
extract complex image features, b) efficient implementation on
\glspl{gpu} \cite{zhangCvpr17}.

\noindent{\it Drawbacks of \glspl{cnnf}: } a) learning is very time
consuming, from few hours to few days, \revv{b) sensitivity to noise
modeling and need of re-training under different noise, c}{b}) inferior
performance on regular textures with high self-similarity.

There have been some attempts to overcome the above-mentioned
drawbacks of \glspl{cnnf}.  In particular, \cite{WavDnCNN} extends
DnCNN \cite{DnCNN} to a feature-space deep residual learning: an input
image is represented by the four subbands of a one-level Haar wavelet
decomposition, which serve as the input to DnCNN.  Despite an improved
performance and partial overcome of the drawback \revv{c}{b}), the method in
\cite{WavDnCNN} is still inferior to some of the recent \glspl{nlf},
especially when regarding to texture \cite{WNNM}.  Some recent
publications consider a mixture of \glspl{nlf} and \glspl{cnn} to
improve the performance on images containing self-similar patches.
The deep \gls{cnn} architecture \cite{Lefk} combines explicit patch
grouping with a learned potential function and regularization
operator.  The so-called \gls{bmcnn} method \cite{BMCNN} is similar,
using groups of similar patches as inputs to a denoising \gls{cnn}.
The structure of this approach resembles that of \gls{bm3d}, but
replaces the Wiener filtering stage by a \gls{cnn} similar to DnCNN.

In this paper, we introduce a paradigm that combines the advantages of
a \gls{cnnf} and of an \gls{nlf} through a simple iterative modular
framework (Sect.~\ref{sec:framework}).  Unlike \cite{Lefk} and
\cite{BMCNN}, which involve new CNN architectures to leverage
nonlocality, ours is a plug-in approach that uses any generic
\gls{nlf} with any generic \gls{cnnf}, without need of retraining.
Furthermore, we present an instance of the proposed framework, called
\thing ~(Sect.~\ref{sec:NLprior}), that enables a nonlocal
self-similarity prior by means of group-wise filtering. Experiments
(Sect.~\ref{sec:experiments}) carried out on several image datasets
demonstrate \thing's ability to exceed the results obtained by its
respective \gls{cnnf} and \gls{nlf} components and achieve
state-of-the-art results in image denoising.

\section{The proposed framework}
\label{sec:framework}

\begin{table*}[!t]
  \centering
  \caption{PSNR ($\mathrm{dB}$) performance of the proposed NN3D vs competitive
    state-of-the-art methods. Italic results in the baseline
    \wavdncnn\ for $\sigma\!=\!75$ indicate scaling of the noisy input
    by the factor 50/75 to match the model trained for
    $\sigma\!=\!50$.}
  \label{tab:performance_psnr}

\pgfplotstableread{data/psnr_set12.csv}\psnrsettwelve
\pgfplotstableread{data/psnr_bsd68.csv}\psnrbsdsixtyeight
\pgfplotstableread{data/psnr_urban100.csv}\psnrurbanhundred

\pgfplotstablevertcat{\fulltable}{\psnrsettwelve}
\pgfplotstablevertcat{\fulltable}{\psnrbsdsixtyeight}
\pgfplotstablevertcat{\fulltable}{\psnrurbanhundred}

\pgfplotstabletypeset[
	create on use/dataset/.style={create col/set list={Set12,,,BSD68,,,Urban100,,}},
	columns/dataset/.style={
		string type,
		column name=Dataset,
		assign cell content/.code={
			\pgfmathparse{int(Mod(\pgfplotstablerow, 3))}%
			\ifnum\pgfmathresult=0%
				\pgfkeyssetvalue{/pgfplots/table/@cell content}
				{\multirow{3}{*}{\emph{##1}}}%
			\else
				\pgfkeyssetvalue{/pgfplots/table/@cell content}{}%
			\fi
		},
	},
    every head row/.style={
    output empty row,
    before row={%
    \toprule
    Dataset & $\sigma$ & BM3D & \multicolumn{1}{c}{DnCNN \cite{DnCNN}} & \multicolumn{1}{c}{\thingcombo{DnCNN}} & \multicolumn{1}{c}{FFDNet \cite{FFDnet}} &  \multicolumn{1}{c}{\thingcombo{FFDNet}} & \multicolumn{1}{c}{\wavdncnn\ \cite{WavDnCNN}} & \multicolumn{1}{c}{\thingcombo{\wavdncnn}}\\
    },
    after row=\midrule,
    },
	every nth row={3}{before row=\midrule},
    columns/sigma/.append style={precision=0},
    every row 2 column wdncnn/.style={postproc cell content/.append style={@cell content/.add={$\it}{$},}},
    every row 5 column wdncnn/.style={postproc cell content/.append style={@cell content/.add={$\it}{$},}},
    every row 8 column wdncnn/.style={postproc cell content/.append style={@cell content/.add={$\it}{$},}},
	columns={dataset, sigma, bm3d, dncnn, cnnfbmdncnn,ffdnet, cnnfbmffdnet, wdncnn, cnnfbmwdncnn},
]{\fulltable}

\end{table*}

We adopt the usual \gls{awgn} observation model:
\begin{equation}
  z=y+\eta\,, \qquad \eta\left(\cdot\right)\sim\mathcal{N}\!\left(0,\sigma^2\right).
\end{equation}
The goal is to estimate the unknown noise-free image $y$ from the
noisy image $z$ and known noise standard deviation $\sigma$.

In the proposed framework, which is summarized by the flowchart in
Fig.~\ref{flowChart} and detailed in Algorithm~\ref{alg:main}, the
noisy image is iteratively filtered by cascaded \gls{cnnf} for
\gls{awgn} removal and \gls{nlf} for enforcing nonlocal
self-similarity.

The rationale of the proposed iterative approach can be explained as
follows. The \gls{cnnf} is biased by the learned mapping of the local
features, while the \gls{nlf} is biased towards nonlocal
self-similarity.  When operating at high noise levels, the local
nature of the \gls{cnn} coupled with the training on external
examples, often leads to \emph{hallucination}, i.e. the introduction
of patterns that do not exist in the original signal $y$. In these
circumstances, the \gls{nlf} becomes especially important as it can
smooth out hallucinations that fail to meet the self-similarity prior,
as we show in Sect.~\ref{sec:experiments}.  \rev{We achieve this by guiding
directly from $z$ the nonlocal self-similarity prior enforced by the
NLF, so that is not correlated with the CNNF.}{}  While on
the one hand the \gls{nlf} can attenuate severe localized artifacts,
on the other hand it might also introduce excessive spatial smoothing.
The proposed framework counteracts this unwanted smoothing by
proceeding iteratively.  Specifically, at each iteration $k$, the
input to the filter cascade is a convex combination
$\bar{z}_k\!=\!\lambda_k z \!+\! (1\! -\! \lambda_k) \hat{y}_{k-1}$ of
the original input $z$ and of the previous estimate $\hat{y}_{k-1}$.
Hence, a fraction of the noise has to be attenuated by the \gls{cnnf},
and thus the \gls{nlf} requires weaker regularization, i.e. a smaller
$\tau_k$. The step parameter $\lambda_k$ controls the rate of progress
of the iterative procedure. Thus, both $\lambda_k$ and $\tau_k$ are
positive and monotonically decreasing with $k$; $\lambda_1\!=\! 1$ as
$\hat{y}_{0}$ is undefined.

\begin{algorithm}[t]
  \caption{Proposed framework}
\label{alg:main}
  \begin{algorithmic}[1]
    \REQUIRE{$z$} \COMMENT{noisy signal}
    \REQUIRE{$\sigma$} \COMMENT{noise standard deviation}%
    \REQUIRE{$K$} \COMMENT{number of iterations}%
    \REQUIRE{$\lambda_k$, $k=1,\dots,K$} \COMMENT{iteration steps}%
    \REQUIRE{$\tau_k,$ $k=1,\dots,K$} {\COMMENT{\gls{nlf} thresholds}}%
    \FOR{$k=1$ \TO $K$}%
    \STATE $\bar{z}_k = \lambda_k z + (1 - \lambda_k) \hat{y}_{k-1}$ \COMMENT{convex combination}%
    \STATE $\tilde{y}_k$ = CNNF$(\bar{z}_k, \lambda_k\sigma)$\label{CNNFalgLine} \COMMENT{CNN-based filter}%
    \STATE $\hat{y}_k$ = NLF$(\tilde{y}_k, \tau_k)$ \COMMENT{Nonlocal filter}%
    \ENDFOR%
    \RETURN $\hat{y}_K$ \COMMENT{return final estimate}%
  \end{algorithmic}
\end{algorithm}

\section{Nonlocal self-similarity prior \& NN3D}
\label{sec:NLprior}

Here we introduce an instance of the above general framework. We call
it \thing, for cascaded Neural Network and 3D collaborative filter,
and it enforces the nonlocal self-similarity prior through the
following two distinct phases:

\noindent \emph{\ A.~Block matching}: identify groups of similar
patches in $z$;

\noindent \emph{\ B.~\gls{nlf} filtering}: shrink spectra of groups
extracted from $\tilde{y}_k$.

\noindent \gls{bm} is executed only once, while the \gls{nlf}
filtering is executed at every iteration.

\subsection{Block matching}
\label{sec:BM}

\gls{bm} is the process by which groups of similar blocks are
identified and it is a crucial step of many \glspl{nlf}
\cite{NL-Bayes, BM3D-TIP, BM3DSAPCA, WNNM}.
\rev{Artifacts in $\tilde{y}_k$ may}{Noise} \rev{severely}{negatively} impact\rev{}{s} the \gls{bm}; therefore, in
\rev{our framework}{\thing}, we do the \gls{bm} on \rev{$z$. In particular,
\thing~performs block matching on}{}\rev{ a filtered}{an} estimate \rev{produced by applying BM3D on $z$}{of $y$, in particular, for convenience, we use the output $\tilde{y}_1$ of the CNNF}:
the result is a look-up
table of group coordinates $S\!=\!\left\{S_1,\dots,S_N\right\}$. Each
$S_j$ contains the coordinates of $N_2$ mutually similar blocks of
size $N_1 \! \times \! N_1$; $S$ is built in such a way that each
pixel in the image is covered by at least one block.
\rev{By obtaining the
group coordinates directly from $z$, we prevent the iterative
procedure from reinforcing artifacts introduced by the CNNF.}{}

\subsection{\gls{nlf} filtering}
\label{ssectFiltering}
 	
Nonlocal self-similarity is enforced by group-wise processing of
$\tilde{y}_k$ based on the look-up table $S$.  Specifically, at each
iteration $k$ and for each $S_j\!\in\!S$, a group
$\mathbf{\tilde{g}}_k^{\left(j\right)}$ is the 3D array of size
$N_1 \!\times\! N_1 \!\times\! N_2$ formed by stacking the blocks
extracted from $\tilde{y}_k$ at the coordinates specified by $S_j$.

While in principle $\mathbf{\tilde{g}}_k^{\left(j\right)}$ could be
filtered by 3D-transform domain shrinkage (akin to \gls{bm3d}), here
the \gls{cnnf} has already operated spatial smoothing and therefore we
perform shrinkage only along the third dimension of the group, with
respect to a 1D transform $\mathcal{T}_{1\text{D}}$ of length
$N_2$. The filtered group is
\begin{equation}\label{eq:NLFshrink}
  \mathbf{\hat{g}}^{\left(j\right)}_k  = 
  \mathcal{T}^{-1}_{1\text{D}}\left(\Upsilon \! \left(\mathcal{T}_{1\text{D}}\big(\mathbf{\tilde{g}}_k^{\left(j\right)}\big), \tau_k\right)\right)\, , 
\end{equation}
where $\Upsilon$ is the shrinkage operator.  Here we define $\Upsilon$
as
\begin{equation}\label{eq:shrinkFun}
  \Upsilon\left(\qWiener, \tau\right)  = 
  \qWiener\,\frac{\qWiener^2}{\,\, \qWiener^2 + \tau^2\,}\, ,
\end{equation}
where the threshold $\tau$ controls the regularization strength with
respect to the nonlocal similarity; its value depends on the magnitude
of the deformations introduced by the \gls{cnnf} as well as on
$\sigma$. Overall, the filtering \eqref{eq:NLFshrink} attenuates
localised deformations that fail to meet the self-similarity
constraint.

The image estimate $\hat{y}_k$ is obtained by returning the block
estimates from the filtered group
$\hat{\mathbf{g}}_k^{\left(j\right)}$ to their original locations
$S_j$, where they are aggregated with group-wise weights reciprocal to
the energy of the shrinkage factor in (\ref{eq:shrinkFun}):
\begin{equation}\label{eq:weight}\notag
w_k^{\left(j\right)} = 
\left\| \frac{\mathcal{T}_{1\text{D}}\big(\mathbf{\tilde{g}}_k^{\left(j\right)}\big)^2}{\,\mathcal{T}_{1\text{D}}\big(\mathbf{\tilde{g}}_k^{\left(j\right)}\big)^2+\tau^2\,}\right\|^{-2}_2\,. 
\end{equation}

The operator $\Upsilon$ can be interpreted as a smoothed firm
thresholding \cite{gao1997waveshrink}, or as a Wiener filter that uses
$\qWiener$ as both the noisy and noiseless coefficient under zero-mean
noise with variance $\tau^2$. Thus, the overall procedure is similar
to the Wiener filter in Similarity Domain~\cite{cruz_2017_single} but
uses
$\mathcal{T}_{1\text{D}}\big(\mathbf{\tilde{g}}_k^{\left(j\right)}\big)$
both as the pilot and as the noisy coefficient to be shrunk.

\section{Experiments}
\label{sec:experiments}

\begin{figure*}[!t]
  \centering
  \showdetailall{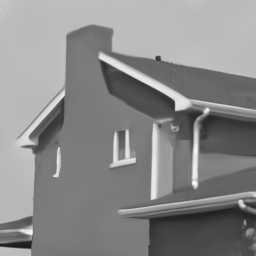}{2.25cm,2.48cm}{.194\textwidth}{0.194\textwidth}{2.05cm}{3.5}{1mm}
  \vspace*{-2.5mm}
  
  \showdetailall{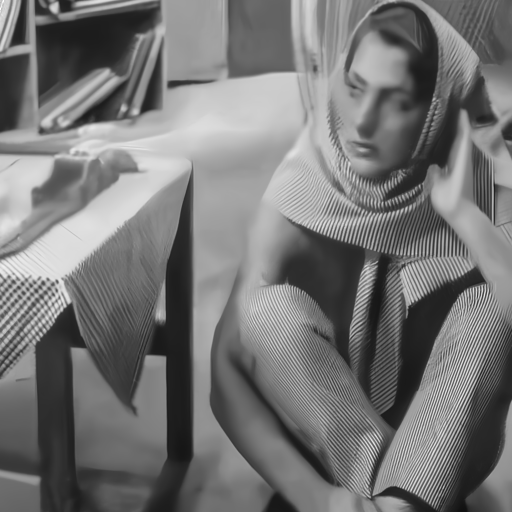}{2.05cm,1.0cm}{.194\textwidth}{.194\textwidth}{2.05cm}{7}{1mm}
  \vspace*{-2.5mm}
  
  \showdetailall{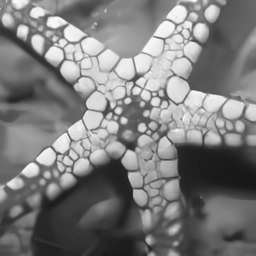}{2.35cm,1.80cm}{.194\textwidth}{.194\textwidth}{2.05cm}{3.5}{1mm}
  \vspace*{-4mm}
  
  \caption{House, Barbara, and Starfish, from \emph{Set12}, with $\sigma\!=\!50$, processed by different methods with respective PSNR.}
  \label{fig:results}
\end{figure*}

We evaluated \thing\ over several datasets: 'Set12', 'BSD68' and
'Urban100'~\cite{datasets}.  We have chosen three state-of-the-art
\gls{cnnf} for AWGN removal: DnCNN~\cite{DnCNN},
\wavdncnn~\cite{WavDnCNN} and FFDNet~\cite{FFDnet}, both for the
comparison and as the \gls{cnnf} module of \thing, where they are used
without any modification or re-training.  For all experiments, our
implementation of \thing\ uses ${K\!=\!2}$, $\lambda_k\!=\!k^{-1}$,
$\tau_k\!=\!\frac{1}{4}\sigma\lambda_k$, $N_1\!=\!10$, $N_2\!=\!32$,
and $\mathcal{T}_{1\text{D}}\!$ is the Haar wavelet transform.

Because the noise variance in $\bar{z}_k$ is approximately
$\lambda^2_k \sigma^2$, the \gls{cnnf} has to be able to operate at
different noise levels. However, \revv{most}{some} \glspl{cnnf} (see, e.g.,
\cite{TNRD}, \cite{DnCNN}) entail models trained for specific noise
standard deviation values
$\cnnTrainStdSet\!=\!\left\{\sigma_1,\dots,\sigma_M\right\}$.  Of the
three \glspl{cnnf} under analysis, \revv{only}{} FFDNet was trained for noise
of arbitrary strength: $\cnnTrainStdSet\!=\!\left[0,75\right]$\revv{. The}{, whereas the}
other two were trained for \revv{very }specific noise levels:
$\cnnTrainStdSet\!=\!\left\{5\!:\!5\!:\!75\right\}$ for DnCNN, and
$\Sigma\!=\!\left\{15,30,50\right\}$ for \wavdncnn.  We thus filter
$\bar{z}_k$ upon scaling it by a factor $\alpha_k$, which is selected
so that the standard deviation of $\alpha_k\bar{z}_k$,
i.e. $\alpha_k\lambda_k\sigma$, matches one of the standard-deviation
levels $\Sigma$ the \gls{cnn} has been trained for. In practice
$\alpha_k$ should not deviate too much from $1$, otherwise
$\alpha_{k} \bar{z}_k$ will not fit the dynamic range of the training
images, and, when $\alpha_k\!>\!1$, $\alpha_{k} \bar{z}_k$ may even
exceed the range adopted for processing, negatively impacting the
performance of the overall scheme. Hence we set
$\alpha_k = \lambda_k^{-1}\sigma^{-1}{\mybreve{\cnnTrainStd}_k}$,
where
$\mybreve{\cnnTrainStd}_k=\max \left\{\left\{\cnnTrainStd \! \in
    \!\smash{\Sigma}\,|\, \cnnTrainStd\!\leq
    \!\lambda_k\sigma\right\}\cup\min \Sigma\right\}$ and Line
\ref{CNNFalgLine} of Algorithm~\ref{alg:main} becomes
$\tilde{y}_k\!=\!\alpha_k^{-1}\text{CNNF}(\alpha_k\bar{z}_k,
\mybreve{\cnnTrainStd}_k)$.

Table~\ref{tab:performance_psnr} reports \gls{psnr} results of the
evaluated methods. For stronger noise, \thing\ noticeably outperforms
the state-of-the-art methods.  Maximal gain over the adopted
\gls{cnnf} modules is achieved on images having strong
self-similarity, such as those from the set 'Urban100'.

In Fig.~\ref{fig:results} one can see that \gls{bm3d} outperforms
\wavdncnn\ on 'Barbara', while \wavdncnn\ outperforms \gls{bm3d} on
'House' and 'Starfish'. However, in all three instances, our approach
outperforms \gls{bm3d} and \wavdncnn, numerically and visually.  In
'Starfish', \thing\ is able to produce sharper results than both
\wavdncnn\ and \gls{bm3d}, showing that the proposed filter cascade is
able to significantly exceed its building blocks.  The results in the
table are also better than those obtained by other proposals combining
nonlocal processing and CNNs~\cite{Lefk}, \cite{BMCNN}. In
\cite{Lefk}, $26.07$ dB is reported when processing 'BSD68' with
$\sigma \!=\!50$, a value which is below all of our tested
combinations. As for \cite{BMCNN}, processing 'House' and 'Barbara'
with $\sigma \!=\! 50$ results in $30.25$ dB and $26.84$ dB
respectively, values which are also significantly below our best
results reported in Fig.~\ref{fig:results}.

\revv{}{As mentioned in Section \ref{sec:BM}, the effectiveness of \gls{bm} can be seriously compromised if this were applied on a noisy image; in particular, if \gls{bm} is applied directly on $z$ instead of on $\tilde{y}_1$, the average PSNR loss over the experiments reported in Table \ref{tab:performance_psnr}  ranges from approximately $0.1$ dB when $\sigma = 30$ to $0.6$ dB when $\sigma = 75$. Nonetheless, the}
\rev{}{\revv{The}{} proposed framework is generic enough that \gls{bm} can \revv{also }{}be performed on the estimate of $y$ provided by another filter\revv{, such as BM3D. We}{. For instance, we} observed that the results obtained \revv{with such a setup, are, in terms of both average PSNR and visual quality, equivalent to the ones presented in this section.}{by performing \gls{bm} on an estimate provided by \gls{bm3d} are equivalent to the ones presented in this section, 
both visually and quantitatively (within $0.05$ dB from the average PSNRs reported in Table \ref{tab:performance_psnr}).} 
}

\rev{}{These experiments were conducted on a computer running Ubuntu 16.04 LTS, equipped with an AMD Ryzen Threadripper 1950X CPU and an Asus GeForce GTX 1080 Ti 11GB Turbo GPU. When processing a 256$\times$256 image, we observed the following average execution times\revv{,}{} for a single call of each module.
Running single threaded on CPU: \revv{BM3D 0.39~s; }{}\gls{bm} 0.10~s; NLF 0.11~s. 
Running on CPU $+$ GPU: DnCNN 0.06~s; FFDNet 0.04~s; WDnCNN 0.42~s.
A complete execution of NN3D ($K\!=\!2$) requires the execution of the following modules: CNNF, BM, NLF, CNNF, NLF.
}
The \thing\ code used for these experiments can be downloaded from
\url{http://www.cs.tut.fi/sgn/imaging/nn3d}.


\section{Discussion and Conclusion}
Even though the receptive fields (a.k.a. field of view) of the
analyzed \glspl{cnn} are in principle large enough to capture nonlocal
self-similarity ($35\!\times\!35$ for DnCNN,
\rev{$31\!\times\!31$}{$62\!\times\!62$} for FFDNet,
\rev{$41\!\times\!41$}{$82\!\times\!82$} for W\rev{av}{}DnCNN, vs.
$39 \!\times\! 39$ search neighborhood of the \gls{bm} in
\thing), their distribution of impact follows a Gaussian distribution,
meaning that the \emph{effective} receptive field of this type of deep
\glspl{cnn} is significantly narrower \cite{luo_2016_understanding}.
Pixels at the center of the field contribute much more than those
towards the periphery. Furthermore, the effective receptive field
grows with the square root of the depth of the network.  This factor
alone explains why these networks are strongly biased towards local
features and why the inclusion of a nonlocal element results in a
dramatic performance improvement.

We showed that the proposed filter cascade is able to boost the
performance of current state-of-the-art \gls{cnnf}.  This is achieved
in part by mitigating hallucinations introduced by \gls{cnnf},
especially when dealing with higher noise variance.  This type of
artifacts violate the self-similarity prior and are therefore
attenuated by the \gls{nlf}.  Overall, the proposed approach yields
cleaner images with much sharper reconstruction of details.

Finally, the modular nature of the proposed framework allows for the
use of whichever \gls{nlf} best integrates with the \gls{cnnf},
processing environment, and data at hand.

\clearpage

\bibliographystyle{IEEEtranN}
\bibliography{bibtex/bib/IEEEabrv,bibliography}

\end{document}